\documentclass[12pt]{article}             
                                   
\usepackage[T1]{fontenc}                  
\newcommand{\3}{\ss}
\newcommand{\absatz}{\vspace{2ex}\noindent}
\newcommand{\bib}[1]{\bibitem{#1}}

\def\journal#1#2#3#4{{#1} {\bf{#2}}, #3 (#4)}

\def\APNY{\emph{Ann.\ Phys.\ } (NY)}
 
\def\IJMP{\emph{Int.\ J.\ Mod.\ Phys.\ }}
\def\LMP{\emph{Lett.\ Math.\ Phys.\ }}

\def\NP{\emph{Nucl.\ Phys.\ }}
\def\PL{\emph{Phys.\  Lett.\ }}

\def\PR{\emph{Phys.\ Rev.\ }}
\def\PS{\emph{Phys.\ Scr.\ }}
\def\RMP{\emph{Rev.\ Mod.\ Phys.\ }}

\newcommand{\dis}{\displaystyle}
\newcommand{\half}{\frac{1}{2}}
\newcommand{\e}{\mathrm{e}}
\newcommand{\ii}{\mathrm{i}}
\newcommand{\dd}{\mathrm{d}}

\newcommand{\Ce}{\mathcal{C}}

\newcommand{\xv}{\vec{x}}
\newcommand{\xvp}{\vec{x}_{\perp}}
\newcommand{\ix}{(\xv)}
\newcommand{\ixp}{(\xv_{\perp})}
\newcommand{\dex}[1][]{\dd x_{#1}\;}
\newcommand{\dezweix}{\dd^{2}\:\!\! x\;}
\newcommand{\dezweixp}{\dd^{2}\:\!\! x_{\perp}\;}
\newcommand{\dedreix}{\dd^{3}\:\!\! x\;}
\newcommand{\dedx}{\dd^{d}\! x\;}

\newcommand{\Ut}{\tilde{U}}
\newcommand{\Av}{\vec{A}}
\newcommand{\Ap}{\vec{A}_{\perp}}
\newcommand{\Piv}{\vec{\Pi}}

\newcommand{\calH}{\mathcal{H}}
\newcommand{\calJ}{\mathcal{J}}
\newcommand{\calM}{\mathcal{M}}
\newcommand{\calO}{\mathcal{O}}
\newcommand{\calS}{\mathcal{S}}
\newcommand{\calU}{\mathcal{U}}

\newcommand{\rtimes}{\mathbin{\times\mkern-3mu%
\raisebox{0.9pt}{\hbox{\vrule width0.2pt height4.5pt depth0.9pt}}}\,}

\def\One{{\mathchoice {\rm 1\mskip-4mu l} {\rm 1\mskip-4mu l}
{\rm 1\mskip-4.5mu l} {\rm 1\mskip-5mu l}}}

\def\CC{{\mathchoice {\setbox0=\hbox{$\displaystyle\rm C$}\hbox{\hbox
to0pt{\kern0.4\wd0\vrule height0.9\ht0\hss}\box0}}
{\setbox0=\hbox{$\textstyle\rm C$}\hbox{\hbox
to0pt{\kern0.4\wd0\vrule height0.9\ht0\hss}\box0}}
{\setbox0=\hbox{$\scriptstyle\rm C$}\hbox{\hbox
to0pt{\kern0.4\wd0\vrule height0.9\ht0\hss}\box0}}
{\setbox0=\hbox{$\scriptscriptstyle C$}\hbox{\hbox
to0pt{\kern0.4\wd0\vrule height0.9\ht0\hss}\box0}}}}

\def\ZZ{{\mathchoice {\hbox{$\sf\textstyle Z\kern-0.4em Z$}}
{\hbox{$\sf\textstyle Z\kern-0.4em Z$}}
{\hbox{$\sf\scriptstyle Z\kern-0.3em Z$}}
{\hbox{$\sf\scriptscriptstyle Z\kern-0.2em Z$}}}}

\def\RR{\mathrm{I\!R}} 

\begin{document}

%

\begin{titlepage}
\begin{flushright}
  hep-ph/9709462\\
  FAU-TP3-97/6 \\
  DOE/ER/41014-35-N97\\
  26th August 1997\\
\end{flushright}
\vspace*{1.5cm}
\begin{center}
  
  \LARGE{\textbf{Magnetic Defects Signal Failure \\ of Abelian Projection
      Gauges in QCD}}

\end{center}
\vspace*{1.0cm}
\begin{center}
  \textbf{Harald W. Grie\3hammer\footnote{Email:
      hgrie@phys.washington.edu}}
  
  \vspace*{0.2cm}
  
  \emph{Institut f\"ur Theoretische Physik III, Universit\"at
    Erlangen-N\"{u}rnberg,\\Staudtstra\3e 7, 91058 Erlangen,
    Germany\\ and \\ Nuclear Theory Group,
    Department of Physics, University of Washington,\\ Box 351 560, Seattle, WA
    98195-1560, USA\footnote{Address after 15th September 1997}}
    \vspace*{0.2cm}

\end{center}

\vspace*{2.0cm}


\begin{abstract} 
  Magnetically charged Abelian defects are shown to arise on most compact base
  manifolds and in most Abelian projection gauges. They obey the Dirac
  quantisation condition and give rise to homogeneous magnetic background
  fields. The reasons for their occurrence are global failures of the procedure
  with which gauge covariant operators are diagonalised or their eigenphases
  extracted.  Defects related to the former case are string-like; for the
  latter case they resemble domain walls. Either configuration forms the
  generic case and indicates a failure of gauge fixing as continuity and
  periodicity properties of the functional space are changed. These results are
  first obtained in canonically quantised QCD$_{3+1}$ and path integral
  QCD$_{2+1}$ on the torus for the modified axial gauge which keeps only the
  eigenphases as dynamical variables of the Wilson line in the $x_3$-direction.
  In the end, they are extended to other gauges, dimensions and standard
  manifolds.
\end{abstract}

\vskip 1.0cm
\end{titlepage}

\setcounter{page}{2} \setcounter{footnote}{0} \newpage

%

\section{Introduction}
\label{intro}
\setcounter{equation}{0}

The non-perturbative r\'egime of non-Abelian gauge theories is rich in open
problems such as the nature of the confinement mechanism and chiral symmetry
breaking, the occurrence of condensates and effective masses, etc. As a
formulation in terms of unconstrained, ``physical'' variables is the easiest
way to render gauge invariant results in approximations, one eliminates the
redundant variables by gauge fixing. Because all well-defined gauge choices are
in the end equivalent but some will yield results and interpretations faster
than others, one hopes that with an appropriate gauge choice, the relevant
degrees of freedom can be identified more easily, the non-perturbative part of
which may solve the outstanding questions in the low energy r\'egime. Recent
success of the Abelian projection gauges in $3+1$ dimensions
\cite{tHooft2,KSW}, which seem to be a useful device in explaining confinement
by the dual Mei\3ner effect \cite{tHooft4,Mandelstam}, triggers the question
how feasible these gauges are for a Hamiltonian formulation in $3+1$ dimensions
and whether for lattice formulations, subtleties are hidden in gauge fixing on
a torus. In this paper, attention focuses particularly on the modified axial
gauge \cite{Yabuki,LNT} in which -- in contradistinction to the na\"ive axial
gauge $A_3=0$ -- the eigenphases of the Wilson line in $x_3$-direction are kept
as dynamical variables.

The advantages of a compact base manifold, especially of a torus $T^d$ in $d$
dimensions as centred on in this paper, are well known. Most prominently
features the automatic infrared regularisation by allowing for zero modes. This
yields a simple way to describe long range fields, condensates, and non-trivial
background fields. In addition, the solution of differential equations (e.g.\ 
the construction of Green's functions) requires a specification of boundary
conditions. The asymptotic hypothesis usually invoked in the infinite volume
limit corresponds to standing wave boundary conditions and hence effectively
compactifies the base manifold. As the example of the na\"ive axial gauge
shows, this yields subtleties hidden in the behaviour of fields at infinity
whose cure in the infinite volume is not fully understood. The torus is the
only orientable manifold which has as universal covering the Euclidean space,
allowing for a global Cartesian co-ordinate system and for translation
invariance. In the canonical formulation, the absence of curvature guarantees
the absolute convergence of the Hilbert space on the torus to the one of the
continuum theory. The torus is also the preferred manifold for lattice
calculations. Although this article foots in its interpretations on the
canonical formulation of gauge theories in $(d+1)$-dimensional space-time,
especially on tori $T^d$ with $d\le3$ and focus on the real world value $d=3$,
the presentation applies equally well to $d$-dimensional path integral
formulations both in Euclidean and Minkowski space, as the next section will
recall.

\absatz The contents are summarised as follows: Section \ref{gfixing}
recalls the philosophy of gauge fixing to the modified axial gauge. The
magnetically charged configurations found on tori $T^d$ in $d\le3$ dimensions
in Sect.\ \ref{defects} do not have particle character and are related to the
extraction of the eigenphases (Sect.\ \ref{eigenphasedefects}) and to the
diagonalisation (Sect.\ \ref{diagonalisationdefects}) of the Wilson line. They
change the boundary conditions (Sect.\ \ref{summarybc}), and their appearance
indicates a failure of gauge fixing because of incompatibility of local and
global gauge conditions as Sect.\ \ref{consequences} emphasises. It also gives
an interpretation as magnetically charged, Abelian defects and highlights their
physical relevance by investigating their weight in configuration space on the
ground of their energy and connection to the functional space measure. Sections
\ref{othermanifolds} and \ref{othergauges} extend the results to other
manifolds and dimensions as well as to the class of Abelian projection gauges.
A summary and an outlook are added in the final section.

First results have been published in Refs.\ \cite{hgtrento,hgprom,hgparis}.
Lenz et al.\ \cite{LNT} already pointed out that boundary conditions may change
in the modified axial gauge, but their derivation was neither complete nor
rigorous enough, so that defects related to the diagonalisation of the Wilson
line were not accounted for. The analysis presented in this paper adds these
and is on a more formal basis, drawing in Sect.\ \ref{diagonalisationdefects}
techniques from Gross et al.\ \cite{GPY} and from topological considerations.
Further new aspects are the interpretation of defects as non-zero magnetic flux
configurations and as signals of a failure of the gauge fixing procedure for
the vast majority of configurations. The extension to the Abelian projection
gauges is also new.

\section{Gauge Fixing to the Modified Axial Gauge}
\label{gfixing}
\setcounter{equation}{0}

In $SU(N)$ gauge theories on a torus $T^d$, $d\le3$, with length of the edge
$L$, one imposes without loss of generality periodic boundary conditions for
the fundamental fields and for the gauge transformations. Including fermions,
one cannot allow for twisted boundary conditions \cite{tHooft5}.
\begin{equation}
    \label{pbc}
        \Av(\xv) =\Av(\xv+L\vec{e}_i)\;\;,\;\; \psi(\xv) = \psi(\xv+L\vec{e}_i)
\end{equation} 
\begin{equation}
        \label{pbcvt}
        V(\xv) =V(\xv+L\vec{e}_i)  
\end{equation} 

In the canonical formulation, the fundamental variables are the gauge and
fermionic fields $\Av\ix,\;\psi\ix$ and their respective conjugate momenta
$\Piv\ix,\;i\psi^\dagger\ix$. They are strictly speaking operator valued
distributions out of which an operator algebra is formed over the functional
space $\calH$ on the space of functions on the torus. Nonetheless, they are
bounded and continuous for all practical purposes \cite[p.\ 106ff.]{Haag}.
Eq.\ (\ref{pbc}) then states that their matrix elements are single-valued,
i.e.\  periodic and continuous functions on the torus. In the same sense, the
sub-algebra of gauge invariant operators becomes a functional sub-space
$\calH_\mathrm{phys}\subset\calH$. In the following, the distinction between
operators and their matrix elements will be dropped in order to simplify
notation.

In the path integral formalism, the fundamental variables are the fields
$\Av\ix$ and $\psi\ix$. $\xv$ includes time, in contradistinction to the
canonical formalism where time enters only as parameter in the Heisenberg
equations of motion. In $d\not=3$, $x_3:=x_d$ still denotes the direction of
gauge fixing. The direction of gauge fixing can also be taken to be (Euclidean
or Minkowski) time, so that one obtains the modified temporal (Weyl) gauge.
The paths run over all configurations on the torus, i.e.\ over all
configurations obeying (\ref{pbc}) and hence again over the same field space
$\calH$ as above.

In both cases, ``gauge fixing'' corresponds to a co-ordinate transformation in
$\calH$
\begin{equation}
  \label{coordtrafo}
  \Av\ix=\Ut\ix\Big[\Av^\prime\ix +\frac{\ii}{g}\;\vec{\partial}\Big]
  \Ut^{\dagger}\ix
\end{equation}
to a basis splitting explicitly in unconstrained variables $\Av^\prime\ix$
which form a basis for the physical variable space $\calH_\mathrm{phys}$, and
constrained ones, $\Ut\ix$ (and respective conjugate momenta in the canonical
formulation). The unconstrained fields become fundamental and are the only
variables present in the Hamiltonian or Lagrangean.

The modified axial gauge \cite{Yabuki,LNT} has often been chosen since for it,
the splitting into $\Ut\ix$ and $\Av^\prime\ix$ can -- as it seems-- be given
concretely, enabling the construction of the Hamiltonian \cite{LNT}, Lagrangean
and Faddeev-Popov determinant \cite{Yabuki,Reinhardt} in terms of primed
variables. The ``zero mode'' fields $A_3^\prime\ixp$ obey the modified axial
``gauge condition''
\begin{equation} 
  \label{gfix}
  A_3^\prime\ixp\mbox{ diagonal}\;\;,\;\;\partial_3 A_3^\prime\ixp= 0\;\;,
\end{equation}
and must remain relevant degrees of freedom since $A^\prime_3\ixp$ are the
phases of the gauge invariant eigenvalues $\exp \ii gLA_3^\prime\ixp$ of the
Wilson line in $x_3$-direction, describing (perturbatively) physical gluons
moving in the transverse sub-torus, $\xvp\in T^{d-1}_\perp$. This is also
expressed in the fact that the solution to (\ref{coordtrafo}) cannot be given
for $A_3^\prime\ix=0$ since then $\Ut$ would not be periodic in $x_3$-direction
and hence in that case one would leave the functional space $\calH$ of periodic
functionals. Allowing for a colour diagonal zero mode, one finds as
$x_3$-periodic solution to (\ref{coordtrafo}) and (\ref{gfix}) at a point on
$T^{d-1}_\perp$
\begin{equation}
   \label{solnut}
   \dis\Ut\ix= \mathrm{P}\e^{\ii g \int\limits_0^{x_3} \dd y_3\;A_3(\xvp,y_3)}
         \Delta\ixp\; \e^{-\ii gx_3 A_3^\prime\ixp}\;\;,
\end{equation}
where $\Delta\ixp$ diagonalises the Wilson line,
\begin{equation}
   \label{diag}
   \Delta\ixp\;\e^{\ii gLA_3^\prime\ixp}\;\Delta^{\dagger}\ixp=
        \mathrm{P}\!\exp\ii g\int\limits_0^L \dex[3]A_3\ix\;\;. 
\end{equation}
An additional gauge choice necessary to fix the gauge completely will be left
out here as the arguments presented apply equally well without it.

\absatz Eq.\ (\ref{solnut}) is not well defined when two or more of the
eigenvalues of the Wilson line are degenerate at some point $\xv_{\perp\,0}\in
T^{d-1}_\perp$.  Such points in configuration space will play an important
r\^ole in the question with which weight the defect configurations derived in
the next section have to be considered. This discussion is postponed to Sect.\ 
\ref{measureanddefects}.

In addition to this problem, another word of caution is in order here. It is
clear that if no degeneracies occur, (\ref{solnut}) is a \emph{local} solution
to (\ref{coordtrafo}) with the gauge choice (\ref{gfix}), i.e.\ that for a
given $\xvp$, one can expect the eigenvalues to be the only physical variables
of the Wilson line. Mind that in both equations, $\xvp$ is merely a label.
Still, that $\Ut\ix$ as given by (\ref{solnut}) is also a solution
\emph{globally}, i.e.\ that it can be chosen regular and periodic in \emph{all}
directions for (nearly) \emph{all} configurations and simultaneously for all
points on $T^d$ is not self-understood. This is the crucial point in the
following: A co-ordinate transformation in $\calH$, built over the space of
periodic functions, cannot change the periodicity of the field distributions in
any of the directions on $T^d$. For a legitimate gauge choice, $\Av^\prime\ix$
acts solely inside $\calH_\mathrm{phys}\subset\calH$ and hence must be
periodic. If this condition is violated, the co-ordinate transformation is not
well defined and one may miss key properties of the non-perturbative sector of
QCD since boundary conditions are tested by global, i.e.\ infrared sensitive,
observables.

The next section will demonstrate that the \emph{local} gauge fixing condition
(\ref{gfix}) and above defined \emph{global} condition on the fields on the
functional space $\calH$ do indeed not match, as Sect.\
\ref{measureanddefects} shows for \emph{most} of the field configurations in
$\calH_\mathrm{phys}$. A forthcoming publication \cite{hgpub2} will show that
the $(3+1)$-dimensional canonical formulation looses consequently the physics
connected to the vacuum-$\vartheta$-angle.

\section{Occurrence of Magnetic Defects}
\label{defects}
\setcounter{equation}{0}

\subsection{General Considerations}
\label{gencons}

It will now be proven that single-valuedness of the fields acting in $\calH$ is
lost during gauge fixing because of the presence of magnetically charged
defects \cite{hgprom}. Note from (\ref{coordtrafo}) and uniqueness of $\Av\ix$
that $\Av^\prime\ix$ is single-valued only if $\Ut\ix$ is. Although the latter
is by construction periodic and continuous in the $x_3$-direction, it is in
general not in $\xvp$ \cite{LNT,hgprom}.

Single-valuedness on $T^d$ is formulated as follows: Going along an arbitrary
closed path $\Ce$ which starts at a point $\xv\in T^d$ and is parametrised by
$s\in[0;1]$, all operators are unique,
\begin{equation}
   \label{singlebc}
    \Av\ix\Big|_{s=1} = \Av\ix\Big|_{s=0}\;\;,\;\;
    \psi\ix\Big|_{s=1} = \psi\ix\Big|_{s=0}
    \;\;\forall\,\xv\;,\;\Ce\in T^d\;\;,
\end{equation}
and the allowed gauge transformations are the ones single-valued on $T^d$ as
well,
\begin{equation}
   V\ix\Big|_{s=1} =V\ix\Big|_{s=0}\;\;\;\;\forall\,\xv\;,\;\Ce\in T^d\;\;.
\end{equation}
In the following, the arguments will be given for the gauge fields only,
extension to the other fundamental variables $\psi\ix$ (and in the canonical
formulation $\Piv\ix$) being straightforward. Now, (\ref{pbc}/\ref{pbcvt})
follow from the identification of the points $\xv^{(i)}$ and
$\xv^{(i)}+L\vec{e}_i$ on $T^d$. Here, $\xv^{(i)}\,,\,i=1,..,d$, denotes a
point with vanishing $i$th component on the boundary of the $d$-dimensional box
which upon identification of opposite sides becomes the torus.

On the other hand, using (\ref{coordtrafo}), one derives that the primed fields
are not necessarily single-valued,
\begin{equation}
  \label{bcloops}
    \Av^\prime\ix\Big|_{s=1} = u_{\Ce}\ix \Big[\Av^\prime\ix\Big|_{s=0} +
    \frac{\ii}{g}\;\vec{\partial}\Big]u_{\Ce}^{\dag}\ix 
    \;\;\forall\,\xv\;,\;\Ce\in T^d\;\;
\end{equation}
\begin{equation}
  \label{uloops}
  \mbox{with }\; u_{\Ce}\ix:=\Ut^{\dag}(\xv)\Big|_{s=1}\Ut\ix\Big|_{s=0}\;\,,
\end{equation}
and at the boundaries of the box
\begin{equation}
  \label{ui}
    u (\vec{x}^{\,(i)} ):= \Ut^{\dag}(\xv^{\,(i)} + L \vec{e}_i) 
        \;\Ut(\vec{x}^{\,(i)} ) \;\;.
\end{equation}

By construction, $\Ut\ix$ is single-valued in the $x_3$-direction, so that for
loops $\Ce$ which wind around the torus only in the $x_3$-direction,
\begin{equation}
   \label{uce3}
   u_{\Ce}\ix=\One\;\;\forall\;\;\Ce=\{\xv=(\xvp,x_3=sL),\;s\in[0;1]\}.
\end{equation}
In order to decide whether $\Ut\ix$ is single-valued, it hence suffices to
examine the properties of each of its terms in (\ref{solnut}) separately under
loops in the transverse sub-torus $T^{d-1}_\perp$.

Since $A_3\ix$ is single-valued on $T^{d-1}_\perp$,
\begin{equation}
   \label{pexpsing}
    \dis\Big[\mathrm{P}\!\exp\ii g\int\limits_0^{x_3}\dd
        y_3\;A_3(\xvp,y_3)\Big]\Big|_{s=1}=\Big[
    \mathrm{P}\!\exp\ii g\int\limits_0^{x_3}\dd 
        y_3\;A_3(\xvp,y_3)\Big]\Big|_{s=0}
    \;\;\;\;\forall\,\xv\;,\;\Ce \in T^{d-1}_\perp\;\;,
\end{equation}
so that with the first term in (\ref{solnut}), the Wilson line is single-valued
on $T^{d-1}_\perp$. Its eigenvalues are defined by (\ref{diag}) only up to
permutations of the order of the entries in a diagonal matrix,
\begin{equation}
   \e^{\ii gLA^\prime_3\ixp} \to R\; \e^{\ii gLA^\prime_3\ixp} \;R^\dagger\;\;.
\end{equation}
Here, $R$ is one of $N!$ members of the Weyl (reflection) group of $SU(N)$, the
$N$-dimensional representation of the permutation group $\calS_N$ embedded in
$SU(N)$. Therefore, strictly not the eigenvalues of the Wilson line are gauge
invariant but the equivalence class
\begin{equation}
   \label{gaugegroup}
   \Big\{R\; \e^{\ii gLA^\prime_3\ixp} \;R^\dagger\;,\;R\in\calS_N\Big\}\;\;
\end{equation}
consisting of every possible Weyl permutation. Eq.\ (\ref{diag}) determines the
residual gauge group as the normaliser of the Cartan sub-algebra, i.e.\ the
equivalence class of all elements in $SU(N)$ which leave its maximal Abelian
sub-group invariant. This is the group
\begin{equation}
    G^\prime=\Big[U(1)\Big]^{N-1}\rtimes \calS_N\;\;, 
\end{equation} 
the semi-direct product of the centraliser of $[U(1)]^{N-1}$, i.e.\ the maximal
Abelian sub-group itself, and of $\calS_N$.  One therefore finds in general on
a closed loop
\begin{equation}
    \label{weyldefect3dim}
    \dis\e^{\ii gLA^\prime_3\ixp}\Big|_{s=1}=R\;\bigg(\e^{\ii gLA^\prime_3\ixp}
    \bigg) \Big|_{s=0} \;R^\dagger\;\;\mbox{ with some } R\in \calS_N\;\;.
\end{equation}
Residual gauge transformations cannot eliminate the factor $R$ because the
mappings of the endpoints of the closed path $\Ce$ into the normaliser
decompose in topologically distinct sectors, $\Pi_0[[U(1)]^{N-1}\rtimes\calS_N]
= \calS_N$. In the following, this possible ``Weyl symmetry'' defect is
neglected and one assumes that $R=\One$. Still, periodicity of all fields in
$\xvp$ cannot be maintained for the two reasons discussed in the next two
subsections.

\subsection{Defects by Extracting Eigenphases}
\label{eigenphasedefects}

The eigenvalue matrices $g\ixp=\exp \ii gLA_3^\prime\ixp\,\in SU(N)$ are
periodic and continuous, and the rightmost term in (\ref{solnut}), $\exp-\ii
gx_3A^\prime_3\ixp$, describes a curve which interpolates between the unit
element ($x_3=0$) and the eigenvalues of the Wilson line ($x_3=L$). The
definition of this term is based on the introduction of a logarithmic
co-ordinate system on the Lie group \cite[p.\ 64]{HaussnerSchwartz} in which
$\exp \ii gx_3A^\prime_3\ixp$ is expressed as a straight line $\ln
g^{x_3/L}\ixp=\ii gx_3A^\prime_3\ixp$, parametrised by $x_3$. A definition of
the eigenphases $A^\prime_3\ixp$ is hence unavoidable.
The mapping
\begin{equation} 
  \label{mapa3}
  \exp \ii gLA_3^\prime\ixp\;:\;T^{d-1}_\perp\rightarrow \Big[U(1)\Big]^{N-1}
\end{equation}
is not simply connected and decomposes into topologically distinct classes
labelled by winding numbers in diagonal, traceless matrices $m_i\in \ZZ^{N-1}$,
$i=1,..,d-1$. Therefore, the phases $A^\prime_3\ixp$ may be chosen to be
continuous (and differentiable) inside $T^{d-1}$ but will in general lie on
different Riemann sheets on opposite boundary points.  In one transverse
dimension, this reads
\begin{equation}
   \label{a3perpbc2dim}
   A^\prime_3(x_1=L)-A^\prime_3(x_1=0)=\frac{2\pi}{gL}\;m\;\;,
\end{equation}
and in 2 transverse dimensions\footnote{This definition of $m_i$ differs from
  the one given in \cite{LNT} by the $\epsilon$-tensor but is more convenient
  later.},
\begin{equation}
   \label{a3perpbc3dim}
   A^\prime_3(\vec{x}^{\,(i)}_{\perp}+L\vec{e}_{\perp,i})-
   A^\prime_3(\vec{x}^{\,(i)}_{\perp})=\frac{2\pi}{gL}\;\epsilon^{ij3}m_j\;\;.
\end{equation}
Lenz et al.\ \cite{LNT} were the first to discuss this change of boundary
conditions. Here, the topological arguments and the interpretation are
added. Respectively, the winding numbers are given by the traceless, diagonal
matrices with $p$th diagonal entry \cite[p.\ 386]{Nakahara}
\begin{equation}
    m_{p}=\dis\frac{g}{2\pi}\int\dex[3]\dex[1]\partial_1 
        A^\prime_{3,p}(x_1)\:\in\ZZ\;\;\mbox{ on }T^2\;\;,
   \label{windingp2d}
\end{equation}
\begin{equation}
   \label{windingp3d}
   \left.
   \begin{array}{r}
   m_{1,p}=\dis\frac{g}{2\pi}\int\dex[3]\dex[2]\partial_2 
        A^\prime_{3,p}\ixp\:\in\ZZ\\[2ex]
   m_{2,p}=-\dis\frac{g}{2\pi}\int \dex[3]\dex[1]\partial_1 A^\prime_{3,p}\ixp
   \:\in\ZZ
   \end{array} 
   \right\}     \;\;\mbox{ on }T^3.
\end{equation}
Because the original configurations are continuous, and hence also the
eigenvalues, the winding numbers cannot depend on $\xvp$.  The eigenphase
defects will be considered in more detail in Sect.\ \ref{measureanddefects}.

\subsection{Defects by Diagonalisation}
\label{diagonalisationdefects}

The second cause of multi-valuedness of $\Ut\ix$, and hence of the primed
fields, is the diagonalisation matrix $\Delta\ixp$. One assumes in the
following that no degenerate eigenvalues occur on $T^2_\perp$. Eq.\
(\ref{diag}) determines then the diagonalisation matrix $\Delta\ixp$ only up to
right multiplication with an element of the Cartan sub-algebra $[U(1)]^{N-1}$
\cite{GPY,FNP}. Therefore, $\Delta\ixp\in SU(N)/[U(1)]^{N-1}$ lies in the coset
of $\exp\ii gLA^\prime_3\ixp$, and with $\Delta\ixp$ a solution to (\ref{diag})
and no Weyl symmetry defects (\ref{weyldefect3dim}) present, any
\begin{equation}
   \label{deltaprime}
      \Delta^\prime\ixp=\Delta\ixp\;h\ixp\;\;\mbox{ with arbitrary } h\ixp\in 
         \Big[U(1)\Big]^{N-1}
\end{equation}
diagonalises the Wilson line on $T^{d-1}_\perp$ as well. So, $[U(1)]^{N-1}$ is
the residual gauge group of this gauge. As all the eigenvalues of the Wilson
line are different, $\Delta\ixp$ is single-valued (and hence periodic) in the
coset, but a priori not in $SU(N)$, so that in general
\begin{equation} 
   \label{deltacurve}
    \Delta\ixp\Big|_{s=1}= \Delta\ixp\Big|_{s=0}
    \;h_{\Ce}\ixp\;\;\;\;\forall \xvp\,,\,\Ce\in T^{d-1}_\perp\;\;.
\end{equation}

Turning to $d=3$ and using homotopy arguments \cite[p.\ 116ff.]{Nakahara}, it
will now be proven that $\Delta\ixp$ is -- without loss of generality -- not
single-valued on $T^2_\perp$, but that one can choose it to be continuous with
a prescribed set of boundary conditions.

The diagonalisation matrix constitutes a continuous mapping of the
2-dimensional closed surface $T^2_\perp$ into the coset on which $ \Delta\ixp$
is periodic, see Fig.\ \ref{mappingfigure}. Therefore, the image of the closed
surface $T^2_\perp$ is again a closed surface on the coset. On $SU(N)$, it is
closed only if there exists another mapping of the surface which $\Delta\ixp$
describes in $SU(N)/[U(1)]^{N-1}$ into $SU(N)$ which is continuous. But the
mapping of the coset into $SU(N)$ is not topological: While the first three
homotopy groups of $SU(N)$ are trivial, $\Pi_d[SU(N)]=\One\;d\le2$, one finds
$\Pi_0[SU(N)/[U(1)]^{N-1}]=\Pi_1[SU(N)/[U(1)]^{N-1}]=\One$ but
$\Pi_2[SU(N)/[U(1)]^{N-1}]=\ZZ^{N-1}$. Recall that the second homotopy group is
isomorphic to the kernel of the natural homomorphism of $\Pi_1[
[U(1)]^{N-1}]=\ZZ^{N-1}$ into $\Pi_1[SU(N)]=\One$, e.g.\ \cite[p.\ 
215ff.]{Coleman}. All elements of $\Pi_1[[U(1)]^{N-1}]$ are mapped onto the
identity of $\Pi_1[SU(N)]$. Therefore, a mapping of the coset into the group
cannot be continuous or periodic in general and images of $2$-dimensional
closed surfaces in $SU(N)/[U(1)]^{N-1}$ will not be closed in $SU(N)$.

\begin{figure}[!ht]
\begin{displaymath}
        \begin{array}{ccc}
        \hbox to0pt{\hss$\mathrm{Map}\Big[\Delta\ixp\Big]\;:\;\;\;\;\;\;\;$}
        \hspace*{0.6em}T^2_\perp & 
        \stackrel{\scriptstyle{d=1:\,\mathrm{closed}\atop d=2:\,\mathrm{open}}}
                {\hbox to5em{\rightarrowfill}}                          &
        SU(N)
        \\ 
        \vcenter{\llap{$\scriptstyle{d=1,2:\atop \hspace*{0.1em}
                \mathrm{closed}}$}}\Bigg\downarrow                      & 
        \vcenter{\vskip 8ex}                                            & 
        \Bigg\uparrow\vcenter{\rlap{$\scriptstyle{d=1:\,\mathrm{closed}\atop
                \hspace*{-0.32em}d=2:\,\mathrm{open}}$}}
        \\
        \hspace*{0.4em}S^2                                              &
        \raisebox{-1.9ex}{$
        \stackrel{\textstyle{\hbox to5em{\rightarrowfill}}}
                {\scriptstyle{d=1,2:\atop\mathrm{closed}}}$}            &
        SU(N)/[U(1)]^{N-1}\hspace*{-1.2em}
        \end{array}
\end{displaymath}
\caption{\label{mappingfigure}
  \textsl{The map of the transverse torus onto \protect$SU(N)$ as described in
    the text. Along the arrow is indicated whether the images of closed
    \protect$d$-dimensional surfaces are closed.}}
\end{figure}

If the manifold which is mapped onto the coset were not $T^2_\perp$ but the
two-dimensional sphere, the proof would now be complete since the second
homotopy group of a manifold is just given by the inequivalent classes of the
mappings of $S^2$ into the manifold. In order to show that topologically
non-trivial maps, i.e.\ non-zero winding numbers, exist for the mapping of
$T^2_\perp$ into the coset, one notes that $ \Pi_1[T^2_\perp]= \ZZ^2$, but
$\Pi_1[SU(N)/[U(1)]^{N-1}]=\One$. So, the mapping cannot preserve topology, and
at least two uncontractible loops on $T^2_\perp$ which also cannot be deformed
into each other have to be mapped to the same element of $SU(N)/[U(1)]^{N-1}$,
thus ``removing the holes'' in the torus (cf.\ smash product \cite[p.\ 
451]{Nakahara} showing ``weak homotopic equivalence'' between $S^d$ and $T^d$).
So, one first maps $T^2_\perp\to S^2$ and then $S^2\to SU(N)/[U(1)]^{N-1}$ as
Fig.\ \ref{mappingfigure} indicates. The winding numbers
$\Pi_2[SU(N)/[U(1)]^{N-1}]=\ZZ^{N-1}$ are preserved if the first map preserves
them. But this is of course always possible, an example being the map of the
torus into the complex plane,
\begin{equation}
   z\ixp=\bigg(\e^{\frac{1}{L-x_1}}-\e^{-\frac{1}{x_1}}\bigg)+
        \ii\bigg(\e^{\frac{1}{L-x_2}}-\e^{-\frac{1}{x_2}}\bigg)\;\;\in\CC
        \;\;,\;\;x_i\in\;]0;L[\;\;,
\end{equation}
and from there to the Riemann sphere by stereographic projection.

In order to obtain the winding numbers of the topologically distinct classes,
one modifies a technique from Gross et al.\ \cite{GPY} and considers the
diagonal and traceless matrix $m_3$ whose $p$th entry is given by
\begin{equation}
 \label{windingmdrei}
    m_{3,p}:=\frac{\ii}{2\pi}\int\dezweixp \vec{\partial}_{\perp}\times
       \Big(\Delta^\dagger\ixp\;\vec{\partial}_{\perp} \Delta\ixp\Big)_p\;\;.
\end{equation}
First, one proves complementarily to the above toplogical reasoning that
$\Delta\ixp\in SU(N)$ is in general not single-valued but may be chosen
continuous inside $T^2_\perp$ so that irregularities occur only at the boundary
of the box, yielding further insight into their nature.  Since the coset is
connected ($\Pi_0[SU(N)/[U(1)]^{N-1}]=\One$), as $SU(N)$ is, one can always
find a diagonalisation matrix which is connected and non-singular in $SU(N)$ on
the \emph{maximal chart} of $T^2_\perp$. This is the open square
which upon identification of opposite sides becomes $T^2_\perp$ and has a
boundary topologically equivalent to a circle $S^1$.

It will now be shown that in general, $\Delta\ixp$ is not everywhere periodic
at the boundary of the maximal chart and hence not single-valued on
$T^2_\perp$. Consider points $\xv_{\perp}^{\,(i)}$ and
$\xvp^{\,(i)}-\vec{\varepsilon}_{\perp}\equiv \xvp^{\,(i)}+L
\vec{e}_{\perp\,,i}-\vec{\varepsilon}_{\perp}$ on the boundary of the square.
They are close neighbours on $T^2_\perp$,
$|\vec{\varepsilon}_{\perp}|\to 0$, but lie far apart in the chart. Since in
general from (\ref{deltaprime}) 
\begin{equation}
        \label{deltaperiodh}
   \Delta(\xv_{\perp}^{\,(i)}+L\vec{e}_{\perp\,,i})=\Delta(\xv_{\perp}^{\,(i)})
   \;h(\xv_{\perp}^{\,(i)}) 
   \;\;,\;\; h(\xv_{\perp}^{\,(i)})\in\Big[U(1)\Big]^{N-1} \;\;, 
\end{equation}
and by construction $\Delta\ixp$ varies continuously inside the square, $
h\ixp$ is continuous on its boundary as well. Using now the freedom
(\ref{deltaprime}) in the diagonalisation of the Wilson line (\ref{diag}), one
can multiply $\Delta\ixp$ from the right with a diagonal matrix $H\ixp$ which
is not necessarily periodic as the square has not yet been glued together to
the torus. The new diagonalisation matrix
\begin{equation}
   \Delta^\prime\ixp:=\Delta\ixp H\ixp
\end{equation}
has the periodicity property
\begin{equation}
  \Delta^\prime(\xv_{\perp}^{\,(i)}+L\vec{e}_{\perp\,,i})=\Delta^\prime
        (\xv_{\perp}^{\,(i)})\;H^\dagger(\xv_{\perp}^{\,(i)})\;
        h(\xv_{\perp}^{\,(i)})\;H(\xv_{\perp}^{\,(i)}+L\vec{e}_{\perp\,,i})
\end{equation}
and is still single-valued on the square when $H\ixp$ constitutes a mapping of
the boundary of the square into $[U(1)]^{N-1}$ of winding number zero, i.e.\ 
\begin{equation}
  \label{hacons}
   \int \dezweixp \vec{\partial}_\perp\times \Big(
   H^{\dag}\ixp\;\vec{\partial}_{\perp} H\ixp\Big)=0\;\;.
\end{equation}
The matrix $m_3$ (\ref{windingmdrei}) is invariant under such transformations.
One can make $\Delta^\prime\ixp$ continuous and piecewise periodic on
$T^2_\perp$ by choosing $H\ixp\in[U(1)]^{N-1}$ such that 
\begin{equation}
    H^\dagger(\xv_{\perp}^{\,(i)})\;h(\xv_{\perp}^{\,(i)})\;
        H(\xv_{\perp}^{\,(i)}+L\vec{e}_{\perp\,,i})=\One
\end{equation}
for some points.  Still, periodicity can usually not be achieved for \emph{all}
points on the boundary of the maximal chart simultaneously because no
continuous function exists on $[U(1)]^{N-1}$ which deforms any mapping of the
boundary continuously into the constant one, $\Pi_1[[U(1)]^{N-1}]=\ZZ^{N-1}$.
The matrix $m_3$ measures these winding numbers. So, for a given Wilson line,
periodic boundary conditions can in general only be retrieved if new
singularities of $\Delta^\prime\ixp$ are introduced inside the maximal chart by
violating (\ref{hacons}) and losing continuity. Like the boundary of the
maximal chart, the position of the multi-valuedness is thus arbitrary and
therefore has no physical significance on its own.

If $H\ixp$ had off-diagonal components in a region inside the maximal chart and
were diagonal only at its boundaries, it could be chosen continuous and such
that $\Delta^\prime\ixp$ were periodic on the torus. But this new
diagonalisation matrix would not diagonalise the same Wilson line (\ref{diag}),
being in a different equivalence class.

In order to prove that $m_3$ measures the total Abelian magnetic flux through
the square $T^2_\perp$, consider following \cite{GPY}
\begin{equation} 
   \label{bedrei}
    b^\prime_{3,p}\ix := \partial_1A^\prime_{2,p}\ix-\partial_2
    A^\prime_{1,p}\ix\;\;,
\end{equation} 
which in the Abelian projection philosophy \cite{tHooft2,KSW} is interpreted as
the QED magnetic field strength in a $(2+1)$-dimensional theory in which the
diagonal gluons are taken as ``photons'', and the off-diagonal ones as
``charged'' vector particles. One inverts the gauge fixing transformation
(\ref{coordtrafo}),
\begin{equation}
     \int\dezweixp b^\prime_{3,p}\ix=\int\dezweixp\vec{\partial}_\perp\times
     \Big(\Ut^{\dag}\ix\Big[\Ap\ix+\frac{\ii}{g}\;\vec{\partial}_{\perp}
     \Big]\Ut\ix\Big)_p \;\;,
\end{equation} 
and notes from (\ref{deltaprime}) that for all closed curves
\begin{equation}
        \label{periodop}
   \oint\limits_{\Ce}\dd\vec{s}_{\perp}\cdot\Big(
   \Delta^\dagger\ixp\;{\calO}\;\Delta\ixp\Big)_p=0
\end{equation}
when $\calO$ is single-valued. Therefore, with $\Ap\ix$ and $\e^{\ii
  gx_3A^\prime_3\ixp}$ being unique on the maximal chart, and (\ref{pexpsing}),
the only non-vanishing term is
\begin{equation}
   \label{winding3}
  \int\dezweixp b^\prime_{3,p}\ix=\frac{\ii}{g}\int\dezweixp
   \vec{\partial}_\perp\times\Big(\Delta^\dagger\ixp\;
   \vec{\partial}_{\perp}\Delta\ixp\Big)_p= \frac{2\pi}{g}\;m_{3,p}\;\;,
\end{equation}
showing that the total Abelian magnetic flux through the box is indeed non-zero
and proportional to the winding number (\ref{windingmdrei}) of the mapping
$\Delta\ixp\,:\,T^2_\perp\to SU(N)/[U(1)]^{N-1}$.

\subsection{New Boundary Conditions and Winding Numbers} 
\label{summarybc}

To summarise, the primed, allegedly physical fields can be chosen single-valued
\emph{inside} the maximal chart of $T^d$ (\ref{bcloops}/\ref{uloops}),
\begin{equation}
   \label{uce}
    u_{\Ce}\ix=\One \;\;\mbox{ for all closed paths $\Ce$ inside the box,}
\end{equation}
but then the boundary conditions (\ref{pbc}) are changed to (\ref{bcloops})
\begin{eqnarray}
  \label{newbc}
   &&\Av^\prime(\vec{x}^{\,(i)}+L\vec{e}_{i})= 
  u^{(i)}(\vec{x}^{\,(i)})\Big[\Av^\prime(\vec{x}^{\,(i)})+
      \frac{\ii}{g}\;\vec{\partial}\Big]u^{(i)\dagger}(\vec{x}^{\,(i)})\;\;,\\
 &&\psi^\prime(\vec{x}^{\,(i)}+L\vec{e}_{i})=u^{(i)}(\vec{x}^{\,(i)})\;
        \psi^\prime(\vec{x}^{\,(i)})\;\mbox{ etc.}\nonumber
\end{eqnarray}
One saw in (\ref{ui}/\ref{uce3}/\ref{a3perpbc3dim}/\ref{deltacurve}) that
without loss of generality in $d=3$
\begin{equation}
   \label{uiexpl3d}
   u^{(i)}(\vec{x}^{\,(i)})=\left\{ \begin{array}{ccl}
            \e^{\frac{2\pi \ii}{L}x_3\epsilon^{ij3}m_j} \;h(\vec{x}^{\,(i)})
                    & \mbox{for}& i=1,2 \\
            \One & \mbox{for}& i=3 \end{array}\right.\;\;,
\end{equation}
and in $d=2$ dimensions (\ref{a3perpbc2dim})
\begin{equation} 
    \label{uiexpl2d}
   u^{(i)}(\vec{x}^{\,(i)})=\left\{ \begin{array}{ccl}
            \e^{\frac{2\pi \ii}{L}x_3 m}& \mbox{for}& i=1\\
            \One & \mbox{for}& i=3 \end{array}\right.\;\;.
\end{equation}
The matrix $h(\vec{x}^{\,(i)})\in [U(1)]^{N-1}$ -- necessary only in the former
case -- is determined by (\ref{deltaperiodh}). It can be chosen to be the unit
matrix on all boundary points simultaneously only if (\ref{hacons}) is obeyed,
i.e.\ if the winding number matrix $m_3$ (\ref{windingmdrei}) is zero.

The winding numbers of the respective mappings are obtained by projecting the
diagonal, traceless matrices with integer entries
\begin{equation}
   \label{windingno2d}
      m_{p} = \frac{g}{2\pi}\int \dezweix b_{p}^\prime\ix =
        \frac{\ii}{2\pi}\;\epsilon^{ij}\int \dezweix\partial_i
      \Big(\Ut^{\dagger}\ix\;\partial_j\Ut\ix\Big)_p \;\; \mbox{ for } d=2\;\;,
\end{equation}
\begin{equation} 
  \label{windingno3d}
     m_{i,p} = \frac{g}{2\pi L}\int \dedreix b_{i,p}^\prime\ix =
        \frac{\ii}{2\pi L}\;\epsilon^{ijk}\int \dedreix\partial_j
        \Big(\Ut^{\dagger}\ix\;\partial_k\Ut\ix\Big)_p \;\;\mbox{ for } d=3
\end{equation}
on the $N-1$ independent $U(1)$ sub-algebrae of the Cartan sub-algebra. Recall
(\ref{coordtrafo}/\ref{solnut}) and (\ref{uce3}/\ref{periodop}), and in
addition for the diagonalisation defect (\ref{winding3}), for the eigenphase
defects (\ref{windingp2d}/\ref{windingp3d}).

Here, $b_{p}^\prime=\partial_1 A^\prime_{3,p}-\partial_3 A^\prime_{1,p}$ and
$b_{i,p}^\prime=\epsilon^{ijk}\partial_j A^\prime_{k,p}$ are the QED magnetic
field strengths of the Abelian projected theory. It will become important in
Sect.\ \ref{measureanddefects} that except for $b^\prime_{3,p}$, they coincide
with the $p$th diagonal entry of the gauge fixed chromo-magnetic field
strength. The total magnetic fluxes of the $U(1)$ sub-algebrae
$\Phi_{i,U(1)}\in \frac{4\pi}{g}\ZZ$ obey the Dirac quantisation condition.
They have no relation to 't Hooft's magnetic twist configurations
\cite{tHooft5}: Magnetic defects are artifacts of the gauge chosen and occur
also when fermions are included as one started with strictly periodic boundary
conditions (\ref{pbc}/\ref{pbcvt}).

In the path integral formulation, one of the $d$ directions is interpreted as
time and hence the winding numbers do not only measure total magnetic Abelian
fluxes but also electric ones. If, for example, $x_3\equiv x_0$ is chosen as
time direction, the interpretation of the diagonalisation defect (\ref{bedrei})
as non-zero net magnetic flux remains, while $\vec{m}_\perp$ measures the total
Abelian electric fluxes $\frac{g}{2\pi L^{d-2}}\int \dedx
\vec{E}_{\perp,p}^\prime\ix$.

Gauge transformations cannot change the defect structure: Neglecting the Weyl
symmetry (\ref{weyldefect3dim}), the eigenphases of the Wilson line are (up to
global shifts by $2\pi$) gauge invariant and hence unaffected by gauge
transformations before or after gauge fixing, showing that $\vec{m}_{\perp}$ is
conserved. A gauge transformation $V\ix$ yields by (\ref{pbcvt}/\ref{diag}) and
the transformation properties of the Wilson line a new diagonalisation matrix
\begin{equation}
         \Delta^\prime\ixp=V(\vec{x}^{\,(3)})
         \;\Delta\ixp\;H\ixp\;\;, \;\;H\ixp\in [U(1)]^{N-1}\;\;,
\end{equation}
and as both $V\ix$ (\ref{pbcvt}) and $H\ixp$ (\ref{hacons}) are single-valued
inside $T^d$, the winding number $m_3$ (\ref{windingno2d}/\ref{windingno3d}) is
invariant. In conclusion, different winding number matrices $\vec{m}$ indeed
distinguish physically different configurations and are not Gribov copies of
each other. Each one represents a different gauge orbit. All of them should be
taken into account in the gauge fixed formulation, especially since Sect.\ 
\ref{measureanddefects} will motivate that they do not form a set of zero
measure in the functional space $\calH_\mathrm{phys}$.

\section{Consequences}
\label{consequences}
\setcounter{equation}{0}

\subsection{Interpreting Defects}
\label{interpretation}

As seen above, the position of the singularities in the fields (and in
$\Ut\ix$) occurring for $\vec{m}\not=\vec{0}$ is of no significance\footnote{In
the following, the notation ``$\vec{m}$'' will be chosen also in $d=2$ where
$m$ is a pseudo-scalar.}, and a priori, no physical particle can be attributed
to it. The defect singularities are hence of the same structure as the Dirac
string in QED with magnetic monopoles. On any closed surface about the QED
monopole, the vector potential cannot be single-valued but has to be singular
on at least one point of arbitrary position. The set of all such points forms
the Dirac string.  In the case of the modified axial gauge in three dimensions,
diagonalisation defects are of co-dimension two and occur in general only at
isolated points on $T^2_\perp$. They form uncontractible strings (vortices)
parallel to the $x_3$-axis, as demonstrated in Sect.\
\ref{diagonalisationdefects}. Eigenphase defects are by construction of
co-dimension one and hence are domain walls parallel to the $x_3$-axis which
wind uncontractibly in the $x_2$-direction when $m_1\not=0$ and vice versa. In
both cases, the magnetic flux through the torus is non-zero and lies in the
Cartan sub-algebra, justifying the name ``magnetic Abelian defects'' for such
configurations.

For the diagonalisation defect, this is especially transparent: Its occurrence
is completely analogous to the introduction of a point singularity on each
sphere $S^2$ about a 't Hooft--Polyakov monopole \cite{tHooft1,Polyakov}. While
the R-gauge solution is regular on any sphere around the monopole position, a
gauge transformation to the unitary gauge, aligning the Higgs field along the
$\sigma^3$-direction in the internal space, introduces a Dirac string which
hits every closed surface about the monopole. It is an artifact of the gauge
fixing \cite[p.\ 58ff.]{Rajamarajan} and can be rotated about the monopole at
will. All local, gauge invariant observables remain well-defined and finite on
the string and monopole position.

It is not surprising that the winding numbers $\vec{m}$ seem to be conserved
under evolution of the primed fields. This can most easily be seen from the
fact that $\vec{m}$ can be written in terms of supposedly constrained variables
only, (\ref{windingno2d}/\ref{windingno3d}), and the gauge fixed Hamiltonian or
Lagrangean contains only allegedly unconstrained (primed) variables. This
interplay between presumed ``physical'' and ``unphysical'' variables
demonstrates once more that for all configurations which exhibit defects, the
modified axial gauge is not well defined. As has been emphasised in Sect.\
\ref{gfixing}, the co-ordinate transformation in field space (\ref{coordtrafo})
is legitimate only if the new basis of the functional space consists again of
periodic fields. Recall that it was the same argument, periodicity of the
fields in $x_3$-direction, which forbade the choice of the na\"ive axial gauge
$A_3^\prime=0$ on a compact manifold. Therefore, the boundary conditions
(\ref{newbc}/\ref{uiexpl3d}/\ref{uiexpl2d}) show that the (local) modified
axial gauge choice (\ref{gfix}) and the (global) requirement for
single-valuedness of all fields on $T^d$ are incompatible. Nonetheless, the
gauge choice is legitimate on any contractible region of the transverse
manifold. Therefore, perturbative results in the modified axial gauge will be
unchanged when one does not take the defects into account. This follows already
from the topological considerations of Sect.\ \ref{defects}: Magnetic defects
only occur for configurations for which the Wilson lines on the transverse
manifold spread out ``far away'' from a constant in group space, i.e.\ far away
from a perturbative vacuum. As Lenz and Thies \cite{LT} emphasised, this vacuum
will not be the configuration $\Av^\prime=0$ when a lattice regularisation of
the Jacobian of the co-ordinate transformation (\ref{jac}) is chosen. So,
defects will be important in the long distance, infrared r\'egime, giving rise
to homogeneous background fields.

\subsection{Weight of Defects in Configuration Space}
\label{measureanddefects}

At first sight, the gauge choice (\ref{gfix}) splits the Wilson line defined on
the gauge group at every point on the transverse manifold into the eigenphases
as allegedly physical and the diagonalisation matrices as allegedly unphysical
variables, (\ref{diag}),
\begin{equation} 
        \label{splitting}
        \mathrm{P}\!\exp\ii g\int\limits_0^L \dex[3]A_3\ix\;\;:\;\; SU(N)\sim
        \Big[U(1)\Big]^{N-1}_\mathrm{phys}\times
        \Big[SU(N)/[U(1)]^{N-1}\Big]_\mathrm{unphys}\;\;.
\end{equation}
This seems to reduce the number of physical gluons locally from $N^2-1$ to
$N-1$. As the gauge group is semi-simple, this cannot be true in general, but
it has been argued \cite{Yabuki,Langetc} that it holds for all configurations
except for a set of measure zero in the functional space $\calH_\mathrm{phys}$.
For this set, the Hamiltonian \cite{LNT} (Lagrangean \cite{JLP}) yields in
addition (na\"ively) infinite energy (action). Indeed, the success of the
Abelian projection gauges, which interpret these singularities as magnetic
monopoles, and of the Haaron model \cite{JLP} suggest that neglecting them is
fallacious in $d=4$ dimensions. The following shows the argument to be
premature for $d\le3$ as well, as $(\vec{m}\not=\vec{0})$-configurations are
not even of zero measure but form in contradistinction the \emph{generic} case
in $\calH_\mathrm{phys}$.

First, note that all sets of configurations with given $\vec{m}$ have the same
cardinality after gauge fixing because the winding numbers
(\ref{windingno2d}/\ref{windingno3d}) are additive: Be
$\Av^\prime_{\vec{m}^{(1)}}\ix$ a configuration with winding number
$\vec{m}^{(1)}$. Adding to it all configurations with winding number
$\vec{m}^{(2)}$, one reaches configurations with winding number
\begin{equation}
   \vec{m}[\Av^\prime_{\vec{m}^{(1)}}+\Av^\prime_{\vec{m}^{(2)}}]=
   \vec{m}^{(1)}+\vec{m}^{(2)}\;\;.
\end{equation}
Because of linearity and the group properties of the winding numbers, this
mapping is one-to-one. But besides cardinality, the measure in the functional
space of primed variables and the action or energy of the configurations must
be taken into account to determine the weight of each configuration in the
gauge fixed functional space.

\absatz The decomposition (\ref{splitting}) is obviously not possible at those
points $\xv_{\perp\,0}$ on the transverse torus on which two or more
eigenvalues of the Wilson line coincide. In that case, one may perform the
gauge fixing in the above way for all other points on $T^2_\perp$, and all
integrals are understood to exclude a small region
$\calU_{\varepsilon}(\xv_{\perp\,0})$ of size $\varepsilon>0$, on which the
gauge has to be fixed separately. Otherwise, renormalisability is manifestly
lost because of ultra-locality of the gauge fixing equations
(\ref{gfix}/\ref{solnut}) in $\xvp$ \cite{tHooft2}. The Jacobian of the
co-ordinate transformation in field space from Cartesian (Lie algebra valued)
co-ordinates $A_3$ to curvilinear co-ordinates, i.e.\ to eigenphases
$A^\prime_3$ of group elements, is the infinite product of Haar measures of
$SU(N)$ at every point \cite{LNT,Reinhardt,JLP},
\begin{equation}
   \label{jac}
    \calJ[A^\prime_3]=\prod\limits_{\xvp}\prod\limits_{p>q} \sin^2
        \frac{gL}{2}\;\bigg(A^\prime_{3,p}\ixp-A^\prime_{3,q}\ixp\bigg)\;\;.
\end{equation}
This is also the Faddeev--Popov determinant for the primed variables in the
path integral formalism. Whenever two or more eigenvalues of the Wilson line
coincide, the dimension of the coset space of $\exp \ii gA^\prime_3\ixp$
increases so that at $\xv_{\perp\,0}$, the Jacobian of the co-ordinate
transformation must be zero in order to retrieve a finite integration
measure. Being a continuous product over all $\xvp$, $\calJ[A^\prime_3]$ is
nonetheless not necessarily zero.

Locally, degeneracy of two or more eigenvalues puts at least three conditions
on a field $A_3^\prime\ixp$ depending on $d-1$ dimensions so that in $d\le3$
such configurations seem negligible. But this argument holds only if no global
reasons exist which support their occurrence. Because of the boundary
conditions (\ref{a3perpbc2dim}/\ref{a3perpbc3dim}) and continuity of the fields
$A^\prime_3\ixp$, one sees that any configuration with
$\vec{m}_\perp\not=\vec{0}_\perp$ has to cross zeroes of the local Jacobian on
an even number of $(d-2)$-dimensional uncontractible hyper-surfaces of
$T_\perp^{d-1}$. As the gauge has been fixed only under the condition that none
of the eigenvalues coincide, eigenphase defects show in a way their own
breakdown.

Before a regularisation is invoked, (\ref{jac}) is only formal because it
involves a continuous product over all space points $\xvp$. On the other hand,
assuming a lattice regularisation \cite{LMT}, $A^\prime_{3,p}$ is interpreted
as polar angle in $SU(N)$ and the other Riemann sheets invoked in Sect.\ 
\ref{eigenphasedefects} are inaccessible. In that case, eigenphase defects
would play no r\^ole for \emph{dynamical} reasons. But on the lattice, one
gives up continuity of the fields and topological effects can only be regained
in the continuum limit by the emergence of smooth lattice configurations. The
subtleties are therefore hidden in the continuum limit. Even if the lattice
interpretation holds in the continuum, eigenphase defect configurations can be
of finite measure in the gauge fixed functional space: Although each of them
might have measure zero, above considerations showed that on the torus, the
vast majority of configurations in $\calH_\mathrm{phys}$ will have eigenphase
defects $\vec{m}_\perp\not=\vec{0}_\perp$. It is hence not clear that a
regularised and renormalised continuum version of the Jacobian (\ref{jac}) will
have the same result, and indeed both 't Hooft \cite{tHooft2} and Johnson et
al.\ \cite{JLP} argue against it. One should also keep in mind that the usual
definition of lattice gauge theory, in which the gauge fields live on the links
rather than the sites, avoids the zeroes of the Jacobian by pressing them in
the centres of plaquettes. It remains to be seen how far the measure really
reduces the contribution of $(\vec{m}_\perp\not=\vec{0}_\perp)$-configurations
to the functional integration.

Diagonalisation defects $m_3\not=0$ can be classified only when no degenerate
eigenvalues of the Wilson line are encountered. Above topological
considerations showed that these defects occur for nearly all configurations
$\Av^\prime\ix$. The Jacobian plays no r\^ole at first sight.  But that $m_3$
is conserved is not stringent since its defining equation (\ref{windingmdrei})
is undetermined whenever eigenvalues coincide in time evolution, reminding one
of the indeterminism in the time evolution of a pendulum at turning point as
discussed in \cite{BW,JMR}.

Therefore, neglecting eigenphase or diagonalisation defects, one throws away
without good reason most of the \emph{generic} configurations which -- as seen
at the end of Sect.\ \ref{summarybc} -- represent physically distinct gauge
orbits. Their fate is intimately connected to a proper renormalisation of the
Jacobian of the co-ordinate transformation in field space.

\absatz The third ingredient which determines with which weight a configuration
enters is its energy in the canonical formalism, or its action in the path
integral representation. In the fully quantised theory, both quantities are
infinite before a renormalisation is invoked, so that the following
considerations are semi-classical only.

It is well known that the Hamiltonian in a $(d+1)$-dimensional theory and the
Euclidean action in a $d$-dimensional formulation coincide formally with only
the mass dimensions of the coupling constants being different. Consider now the
minimum energy a configuration with defect $m$ has in the canonical formulation
for $d=2$. The Hamiltonian is bounded from below and reads after gauge fixing
\cite{LNT}
\begin{equation}
     H=\half\int\dezweix\Big[\big(\Piv^{\prime\,a}\ix\big)^2+
     \big(B^{\prime\,a}\ix\big)^2\Big] + H_\mathrm{fermi} +
     H_\mathrm{colour}\;\ge 0 \;\;,
\end{equation}
where $H_\mathrm{fermi}$ is the minimally substituted fermionic standard
Hamiltonian. $H_\mathrm{colour}$ is a non-negative term generated by resolving
Gau\3' law and describes the interaction between static colour charges. These
terms and the contribution from chromo-electric and off-diagonal
chromo-magnetic fields are positive, so that one obtains a lower bound,
\begin{equation}
    H\ge \half\int\dezweix \big(B^{\prime\,a_0}\ix\big)^2 \;\;.
\end{equation}
As observed below (\ref{windingno3d}), $B^{\prime\,a_0}=b^{\prime\,a_0}$. One
decomposes the chromo-magnetic field now into its non-zero and zero modes,
where the latter is given from (\ref{windingno2d}) by
\begin{equation}
   b^\prime_{\mathrm{z.m.},p}=\frac{2\pi}{gL^2}\;m_p\;\;,
\end{equation}
so that the lower bound is
\begin{equation}
   H\ge \frac{4\pi^2}{g^2L^2}\sum\limits_p m_p^2\;\;.
\end{equation}
The Euclidean action in 1+1 dimensions is correspondingly derived as
$S_\mathrm{E}\ge \frac{4\pi^2}{g^2L^2}\sum_p m_p^2$.  In conclusion, eigenphase
defects are energetically disfavoured over $(m=0)$-configurations, but this
effect dies out in the infinite volume limit.

In $d=3$ dimensions,
\begin{equation}
   H\ge \half\int\dedreix \big(\vec{B}^{\prime\,a_0}\ix\big)^2 \;\;,
\end{equation}
and the same arguments apply to the eigenphase defects,
\begin{equation}
   \half\int\dedreix \big(\vec{B}^{\prime\,a_0}_\perp\ix\big)^2\ge
   \frac{4\pi^2}{g^2L}\sum\limits_p \vec{m}_{\perp, p}^2\;\;.
\end{equation}
But since $B^{\prime a_0}_3=b^{\prime a_0}_3+gf^{a_0bc} A^{\prime b}_1
A^{\prime c}_2$, no non-trivial lower bound can be derived for the energy of
diagonalisation defects by this method. Indeed, consider a configuration
\begin{eqnarray}
    &&\Ap\ix=0\;\;,\nonumber\\
    &&A_3\ix=A^\prime_3\Big[\sigma^1 \sin\vartheta \cos\varphi  + \sigma^2
    \sin\vartheta \sin\varphi + \sigma^3 \cos\vartheta\Big]\;\;,
\end{eqnarray}
where $\sigma^a$ are the Pauli matrices. The pseudo-spherical ``angles'' are
$\vartheta$, $\varphi$ curvilinear co-ordinates which parametrise $T^2_\perp$
such that $\vartheta=0$ is some point in the interior and the closed curve at
$\vartheta=\pi$ is the boundary of the box, parametrised by
$\varphi\in[0;2\pi]$. The parameter $A^\prime_3$ is $\xv$-independent, and the
Wilson line of this configuration has eigenvalues $\e^{-\ii gL
  A^\prime_3\sigma^3}$. By choosing $\Piv=0$ as initial condition, the energy
of this configuration is
\begin{equation}
   H=\half\int\dedreix\Big[\big(\partial_1A^a_3\big)^2
   +\big(\partial_2A^a_3\big)^2\Big] =
   \calO\Big(\big(A^\prime_3\big)^2\Big)\;\;,
\end{equation}
and hence can be chosen arbitrarily small by tuning $A^\prime_3$. On the other
hand, diagonalising the Wilson line shows that after gauge fixing one
encounters a diagonalisation defect $m_3=\sigma^3\not=0$ of winding number
one. Therefore and in contradistinction to the eigenphase defects,
diagonalisation defects carry arbitrarily small energy (action) in three
dimensions. The contribution which na\"ively comes from the zero mode QED
magnetic field $b^\prime_3$ is hence cancelled by the interaction between it
and the ``charged'' vector particles $A^{\prime a_1}_\perp\ixp$. The Dirac
string has zero energy density. The same phenomenon can be observed for the 't
Hooft--Polyakov monopole in the unitary gauge \cite[p.\ 58ff.]{Rajamarajan}.

When eigenphase defects occur in $d=4$ (see Sect.\ \ref{othermanifolds}), each
contribution $S_{m_{\perp}}\ge\frac{4\pi^2}{g^2}\sum_p m_{\perp,p}^2$ to the
action is volume independent and -- like the instanton action -- non-analytic
in the coupling constant.

\subsection{Other Manifolds}
\label{othermanifolds} 

As one believes that short range physics should not be sensitive to the choice
of boundary conditions in a large universe, it is important to note how much
the analysis presented relies on the choice of a torus as base manifold. Mind
that for compatibility with the gauge choice (\ref{gfix}), the closed manifold
should split into a product for the modified axial gauge,
$\calM=S^1_{x_3}\times \calM_\perp$. In two dimensions, all closed, orientable
transverse manifolds are topologically equivalent to the circle $S^1$. Since
$\Pi_1[SU(N)/[U(1)]^{N-1}]=\One$, one can without loss of generality choose
$\Delta\ixp$ to be continuous and periodic everywhere, so that the
diagonalisation defect is not present. Eigenphase defects are still encountered
because $\Pi_1[[U(1)]^{N-1}]=\ZZ^{N-1}$. In contradistinction, the
diagonalisation defect in $d=3$ is encountered on both transverse standard
manifolds $T^2$ and $S^2$ (see Fig.\ \ref{mappingfigure}), but only
$\calM_\perp= T^2$ suffers from eigenphase defects and not the sphere as
$\Pi_2[[U(1)]^{N-1}]=\One$.

In higher dimensions, all manifolds with non-trivial fundamental group
$\Pi_1[\calM_\perp]\not=\One$ will have eigenphase defects in the modified
axial gauge, e.g.\ all tori $T^{d-1}_\perp$. For diagonalisation defects, the
criterion is more difficult to formulate, but at least every direct product
manifold which contains factors $T^2$ or $S^2$ will, by embedding above
mappings, suffer from these. Since in $d\ge4$ the presence of instantons
($\Pi_3[SU(N)]=\ZZ$) hampers in general the choice of periodic boundary
conditions (\ref{pbc}), the winding numbers are usually not given by a simple
generalisation of above formulae. Nonetheless, periodic boundary conditions can
be chosen for three of the directions, and for those the winding numbers are
given by obvious generalisations of (\ref{uiexpl3d}/\ref{windingno3d}).

\subsection{Abelian Projection Gauges}
\label{othergauges} 

As the considerations of Sects.\ \ref{defects} and \ref{consequences} based
mostly on the decomposition of the gauge group into equivalence classes and
cosets w.r.t.\ some diagonalised variable (\ref{splitting}) only, all
conclusions can easily be extended to the class of Abelian projection gauges
\cite{tHooft2,KSW}. One only has to test to which dimensions and manifolds they
apply.

The diagonalisation defects of Sect.\ \ref{diagonalisationdefects} relied only
on the existence of a non-trivial mapping of the manifold into the coset of
some diagonal matrix. Therefore, they occur in any Abelian projection gauge
whose diagonalised operator depends on two co-ordinates which label factors
$T^2$ or $S^2$ on a direct product base manifold. Examples are the $F_{12}$-
and maximal Abelian gauges \cite{tHooft2} on $S^2$ and $T^3$. Eigenphase
defects are present only when the diagonalised operator is a member of a
multiply connected space (eg.\ the Cartan sub-algebra). Therefore,
$\vec{m}_\perp=0$ for the $F_{12}$-gauge independent of the base manifold
because the diagonalised Lie algebra element $F^\prime_{12}\ix\in\RR^{N-1}$.
The maximal Abelian gauge exhibits eigenphase defects for any manifold with
non-trivial fundamental group. Both kinds of defects are encountered on tori
considered in lattice simulations in $d\ge2$, and of all standard manifolds in
$d\le4$, only the spheres $S^3$ and $S^4$ are free of any defects.

So, the occurrence especially of diagonalisation defects signals the failure of
many Abelian projection gauges in the finite volume and entails a change of
boundary conditions according to (\ref{newbc}). In general, the Jacobian will
be different from (\ref{jac}), but it will become locally zero when two or more
eigenvalues become degenerate. On the other hand, the success of these gauges
in the explanation of confinement on $(3+1)$-dimensional lattices suggests that
-- albeit axiomatically not legitimate -- the formal failure may be of only
minor consequence to this aspect of the theory. To track down the reason for
this would be an interesting task. It is also especially useful as Abelian
projection gauges are hardly considered in the canonical formulation of
$(3+1)$-dimensional gauge theories. As shown above, the magnetic defects are
not physical particles, having neither mass nor position, but are artifacts of
the gauge chosen. Therefore, they cannot condense, and the 't Hooft--Mandelstam
mechanism for confinement \cite{tHooft2,tHooft4,Mandelstam} cannot be
associated with them. In fact, Lenz et al.\ \cite{LMT} showed that in the
modified axial gauge, gluons are confined in the strong coupling limit due to
the Jacobian (\ref{jac}).

\section{Conclusions}
\label{conclusions}
\setcounter{equation}{0}

In this article, the modified axial gauge condition in $SU(N)$ gauge theories
(\ref{gfix}) has been investigated which keeps as physical variables of the
Wilson line in $x_3$-direction only its eigenphases. The results apply to
canonical and path integral formulations in $d\le3$ dimensions and to most
Abelian projection gauges, as Sects.\ \ref{othermanifolds} and
\ref{othergauges} showed.

On any open, contractible region of a $(d-1)$-dimensional torus $T^{d-1}_\perp$
perpendicular to the Wilson line, this gauge choice allows for a regular
solution of the co-ordinate transformation (\ref{coordtrafo}) in field space
which fixes the gauge under the condition that none of the eigenvalues of the
Wilson line are degenerate. Still, due to ambiguities in the extraction of the
eigenphases $A_3^\prime\ixp$ of the Wilson line and in the definition
(\ref{diag}) of its diagonalising matrix $\Delta\ixp$, compactification to
$T^{d-1}_\perp$ introduces singularities in the gauge fixed variables. They
have been shown to describe Abelian defects with non-zero magnetic net fluxes,
incorporated by closed loops of Dirac strings which wind around the torus for
diagonalisation defects, and by domain walls for eigenphase defects.
Configurations with $\vec{m}\not=\vec{0}$ feel hence a non-zero magnetic
background field (\ref{windingno2d}/\ref{windingno3d}). Topological arguments
demonstrated that their existence is unavoidable, but the defect position is
irrelevant because the whole transverse space had to be mapped into the maximal
Abelian sub-group and its coset respectively, and not only a neighbourhood of
some point. Different defect winding numbers have been shown to label
physically distinct classes of configurations. For the eigenphase defects, this
is already seen from the fact that each of them has a minimum energy or action
proportional to the square of its winding number (Sect.\ 
\ref{measureanddefects}). In contradistinction, the diagonalisation defects can
have infinitesimally small but positive energy (action). Both defects seem to
be topologically and dynamically stable, but this has been discussed to be a
consequence of the non-legitimate gauge choice.

The change of boundary conditions leads to the conclusion that from the formal
point of view, the modified axial gauge is not a legitimate gauge choice since
the structure of the functional space $\calH$ over which the theory is defined
is altered. In Sect.\ \ref{gencons}, the gauge is fixed not only by the gauge
choice (\ref{gfix}), but also by the condition that the underlying functional
space remains unchanged, i.e.\ that the new physical and unphysical fundamental
variables are single-valued (continuous and periodic in all directions) on the
manifold. This is an implicit but important gauge choice made even before the
co-ordinate transformation in field space (\ref{coordtrafo}) was performed. It
has been demonstrated that the gauge fixing transformation $\Ut\ix$ is not
single-valued and hence not a member of the original functional space $\calH$
on the torus. The \emph{local} and \emph{global} gauge conditions cannot be met
simultaneously for the majority of configurations in the physical field space,
so that the gauge condition does not intersect most of the gauge orbits. In
principle, the new field space is obtained from the original one by a procedure
which allows one to determine the precise nature of the singularities and hence
to enlarge the new configuration space accordingly. But because of the problems
with the measure, i.e.\ the Jacobian (\ref{jac}), this seems not to be a
feasible method, as seen in Sect.\ \ref{measureanddefects}. One also would have
to take into account allegedly unphysical variables and dynamical boundary
conditions, as the winding numbers of the defects is closely related to both of
them, (\ref{uiexpl3d}/\ref{uiexpl2d}) and
(\ref{windingno2d}/\ref{windingno3d}).

Eigenphase defects do not depend on the number of dimensions but on the choice
of the base manifold, while diagonalisation defects occur in the modified axial
gauge especially in $d=3$ dimensions but independent of the specific choice of
$\calM$ (Sect.\ \ref{othermanifolds}). Abelian projection gauges suffer from
one or the other defect, except when $S^3$ or $S^4$ are chosen as base manifold
of the diagonalised variable (Sect.\ \ref{othergauges}). The diagonalisation
defect turned out to be especially persistent. It is interesting that only in
three dimensions, non-Abelian gauge theories allow for the existence of large
gauge transformations which give rise to the vacuum--$\vartheta$--angle in the
canonical formulation or to the quantised topological mass of Chern--Simons QCD
in the path integral. A parallel publication \cite{hgpub2} demonstrates that
defects signal the loss of large gauge transformations for the Abelian
projection gauges, albeit the r\^ole of the Jacobian of the co-ordinate
transformation in field space is unclear. On the other hand, Jahn et al.\ 
\cite{JKS} demonstrated that the eigenphase defects in the modified axial
gauge, path integral formulation of QCD$_{1+1}$ are irrelevant because of the
Jacobian (\ref{jac}) and the interpretation of the variables $A^\prime_3\ixp$
as polar angles. In higher dimensions, an ultra-local interpretation of the
Jacobian is not necessarily correct in the continuum theory, and one may
speculate that in the path integral for QCD$_{2+1}$, defects may be related to
giving linear instead of logarithmic confinement in the low temperature phase
($L$ large). For the Hamiltonian version of QCD$_{3+1}$, the Jacobian as
quantum phenomenon may inhibit or enhance classically allowed processes, but
for a proper understanding of its r\^ole in various dimensions, a proper
regularisation and renormalisation of its continuum version is needed.

\section*{Acknowledgments} 
It is a pleasure to acknowledge intense discussions with O.\ Jahn, A.\ C.\ 
Kalloniatis, and F.\ Lenz. This work was supported in part by the
Graduiertenkolleg Erlangen--Regensburg on Strong Interaction Physics of the
Deutsche Forschungsgemeinschaft and by the Bundesministerium f\"ur Forschung
und Technologie under grants 06 ER 747 and 06 ER 809.


\end{document}